\documentclass[sn-mathphys,Numbered]{sn-jnl}% Math and Physical Sciences Reference Style
\usepackage{natbib}
\usepackage{graphicx}%
\usepackage{multirow}%
\usepackage{amsmath,amssymb,amsfonts}%
\usepackage{amsthm}%
\usepackage{mathrsfs}%
\usepackage[title]{appendix}%
\usepackage{xcolor}%
\usepackage{textcomp}%
\usepackage{manyfoot}%
\usepackage{booktabs}%
\usepackage{algorithm}%
\usepackage{algorithmicx}%
\usepackage{algpseudocode}%
\usepackage{listings}%

\usepackage{algorithm}
\usepackage{arevmath}     % For math symbols
\usepackage{algpseudocode}

\theoremstyle{thmstyleone}%
\theoremstyle{thmstyletwo}%

\theoremstyle{thmstylethree}%

\begin{document}

\title[Article Title]{A New Learning Approach for Noise Reduction}


\author*[1]{\fnm{Negin} \sur{Bagherpour}}\email{negin.bagherpour@ut.ac.ir}

\author[2]{\fnm{Abbas} \sur{Mohamadiyan}}\email{mohamadiyan.777@ut.ac.ir }

\affil*[1,2]{\orgdiv{Department of Engineering Sciences}, \orgname{University of Tehran}, \city{Tehran}, \postcode{14155-6619}, \country{Iran}}


\abstract{Noise is a part of data whether the data is from measurement, experiment or ... A few techniques are suggested for noise reduction to improve the data quality in recent years some of  which are based on wavelet, orthogonalization and neural networks. The computational cost of exisiting methods are more than expected and that's why their application in some cases is not beneficial. In this paper, we suggest a low cost techniques based on special linear algebra structures (tridiangonal systems) to improve the signal quality. In this method, we suggest a tridiagonal model for the noise around the most noisy elements. To update the predicted noise, the algorithm is equipped with a learning/feedback approach. The details are described below and based on presented numerical results this algorithm is successful in computing the noise with lower MSE (mean squared error) in computation time specially when the data size is lower than 5000. Our algorithm is used for low- range noise while for high-range noise it is sufficient to use the presented algorithm in hybrid with moving average. The algorithm is implemented in MATLAB 2019b on a computer with Windows 11 having 8GB RAM. It is then tested over many randomly generated experiments. The numerical results confirm the efficiency of presented algorithm in most cases in comparison with exisitng methods.}




\maketitle

\section{Introduction}\label{sec1}
Data analysis is a very common problem in machine learning and signal processing. Assume that a quantity $X$ is measured in $n$ different cases and we need to analyze the provided data. Since the data contains some error, whether it obtained by experiments or direct measurement tools, we need to detect and reduce the noise before starting any analysis. Different noise reduction algorithms are suggested with specific applications for audios or images; see for example \cite{RC1965, Burwen1971}. In recent years wavelet and least squares played an important role in suggested noise reduction algorithms. Chen \cite{Chen2015} presented an algorithm based on noise orthogonalization. He also outlined a novel noise reduction technique by use of reverse least squares and shaping regularization \cite{Chen2015, Chen2016}. Moreover, he developed the first wavelet based algorithm for noise reduction in 2017 \cite{Chen2017}. On the other hand, Huang \cite{Huang2016} provided a singular spectrum analysis for 3D random noise. A few neural network based algorithms are also suggested for noise detection \cite{Mafi2019, Orazaev2023}. There are some complications in solving noise detection problem such as difficult mathematical modeling, high computational cost and high sensitivity to noise quality and size. Learning approaches can solve these issues by following the error trend in consequtive iterations. In each iteration of a learning algorithm, the current noise estimation is evaluated to suggest a proper update. In this paper we provide a new algorithm for detecting and reducing the noise which has benefits over existing methods in some cases. Each iteration of this algorithm consists of three main steps:\\
1) It suggests a tridiagonal model for the signal entries which show more noisy behavior. \\
2) The noise is approximated by solving the tridiagonal model. \\
3) It updates the input signal considering the assumed noise.

\begin{figure}[!h]%[!htbp]
\centering
\includegraphics[scale=0.3]{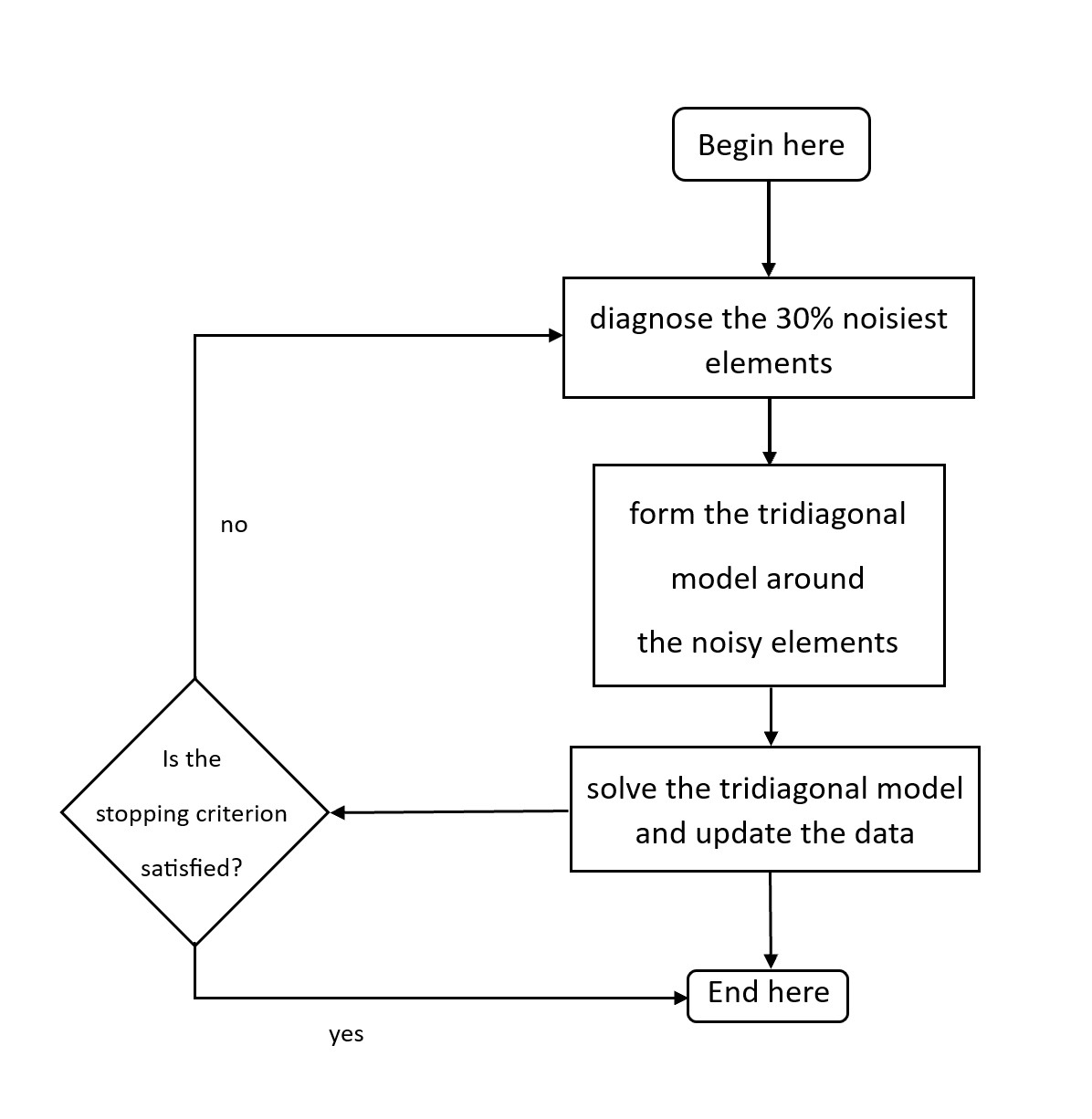}
\caption{Flowchart of the Low-dimension Tridiagonal (LTD) Algorithm}
\label{fig1}
\end{figure}

Our contributions are as follows:\\
1) We outline a two phase noise reduction algorithm which suggests a tridiagonal model to estimate noise, compute an approximated noise-free signal and check the improvement to verify the quality or revise the noise in each iteration.\\
2) The hybrid of regression phase besides the learning phase make the noise reduction process faster.\\
3) The complexity of the proposed algorithm is relatively low.\\
4) We can substitute the tridiagonal model by any proper matrix structure based on signal characteristics. We categorized some special cases in Section \ref{sec3}.

%I beleive we could write s psudocode for it!
\section{Our Algorithm}\label{sec2}
\subsection{Initialization}

The Algorithm starts with guessing noise through simple idea of moving average and normal distribution, These midpoints will be used in subsequent calculations. then goes on with detecting the potential noisy elements from ruined Data and trying to reduce the error. Doing it over and over(in the next step) until the stopping criterion gets satisfied, the stopping criterion would be either predefined by the user or automatically detected from the Main Data itself.

\subsection{Approximation loop}

The algorithm enters a while loop that continues as long as the error E is greater than the specified tolerance (i.e. stopping criterion) and the counter is within the maximum number of iterations that we have predefined.
Inside the loop, the second-order differences of GT are computed, and the maximum value M is determined. This helps identify the elements of interest for the approximation.
The elements from GT that meet the condition $abs(DD)-0.7*M>0$ are selected and stored in gt. These elements will be used in the approximation calculations.
The length of gt is determined and stored as n.
The approximation vector f is initialized as a zero vector with a length of n.
If n is greater than zero, the algorithm proceeds with the approximation calculations.
Within the for loop, a set of input values In is generated, evenly spaced over a range related to the length n. These input values are used to evaluate the PDF PD1 later on.
The PDF values N are computed by evaluating the PDF PD1 using the input values In.
Intermediate arrays and variables are initialized for subsequent calculations.
Random-based calculations are performed to update the elements of f based on the values of N and gt. These calculations involve random coefficients and the proportionate allocation of the PDF values.
A tridiagonal matrix T is constructed based on the calculated values. The diagonal elements are determined by the updated d array, while the off-diagonal elements are determined by the updated mu and rho arrays.
The linear system T * f = N is solved to obtain the updated approximation vector f.
The error E is updated by calculating the norm of the difference between f and gt.
The counter k is incremented by 1 to keep track of the number of iterations.

\subsection{Post-processing and analysis}
Once the while loop finished, the measured values GT are updated by replacing the selected elements based on gt with the corresponding elements from f. This reflects the refined approximation.
The measured values GT are plotted against the exact values Gexact to visualize the approximation and assess the quality of the results.
The mean squared errors (mse1 and mse2) between the exact values Gexact and the measured values Gmeasured are computed and stored. These metrics provide a quantitative assessment of the approximation's accuracy.
The algorithm iterates through the approximation loop, adjusting the approximation vector f based on the calculated PDF values and the selected elements from gt. The goal is to refine the approximation and minimize the error between the measured values and the true values. The process continues until the error falls below the specified tolerance or the maximum number of iterations is reached.

\subsection{Pseudocode}
\begin{algorithm}
\caption{Put your caption here}
\begin{algorithmic}[1]
\Procedure{LDT}{$Nn, \text{mydelta}, kmax$}
    \State Initialize variables and arrays
    \State Generate exact data $G_{\text{exact}}$
    \State Create a probability distribution $PD$ with desired characteristics
    \State Generate noise data $N_{\text{Data}}$ based on the distribution
    \State Add the noise data to the exact data to obtain $GT$
    \State Calculate midpoint values $gm$ from $GT$
    \State Compute mean and standard deviation of differences between $gm$ and $GT$
    \State Initialize error $E$ and measured data $G_{\text{measured}}$
    
    \While{$E > \text{mydelta}$}
        \State Compute second-order differences $DD$ of $GT$
        \State Find the maximum difference $M$ in $DD$
        \State Select elements of $GT$ based on a condition to obtain $gt$
        \State Determine the length $n$ of $gt$
        
        \If{$n > 0$}
            \For{$k = 1$ \textbf{to} $kmax$}
                \State Generate input values $In$
                \State Compute probability density values $N$ based on $PD1$
                \State Initialize intermediate arrays and variables
                
                \State Perform random-based calculations to update $f$
                \State Construct a tridiagonal matrix $T$ based on $d$, $mu$, and $rho$
                \State Solve the linear system $T \cdot f = N$
                \State Update error $E$ based on the difference between $f$ and $gt$
            \EndFor
        \EndIf
        
        \State Update the measured data $G_{\text{measured}}$ based on $gt$ and $f$
    \EndWhile
    
    \State Calculate mean squared errors $mse1$ and $mse2$
    \State \textbf{Return} the final values of $f$, $mse1$, and $mse2$
\EndProcedure
\end{algorithmic}
\end{algorithm}

\section{Some Hints about Convergence}\label{sec3}
\begin{figure}[!h]%[!htbp]
\centering
\includegraphics[scale=0.2]{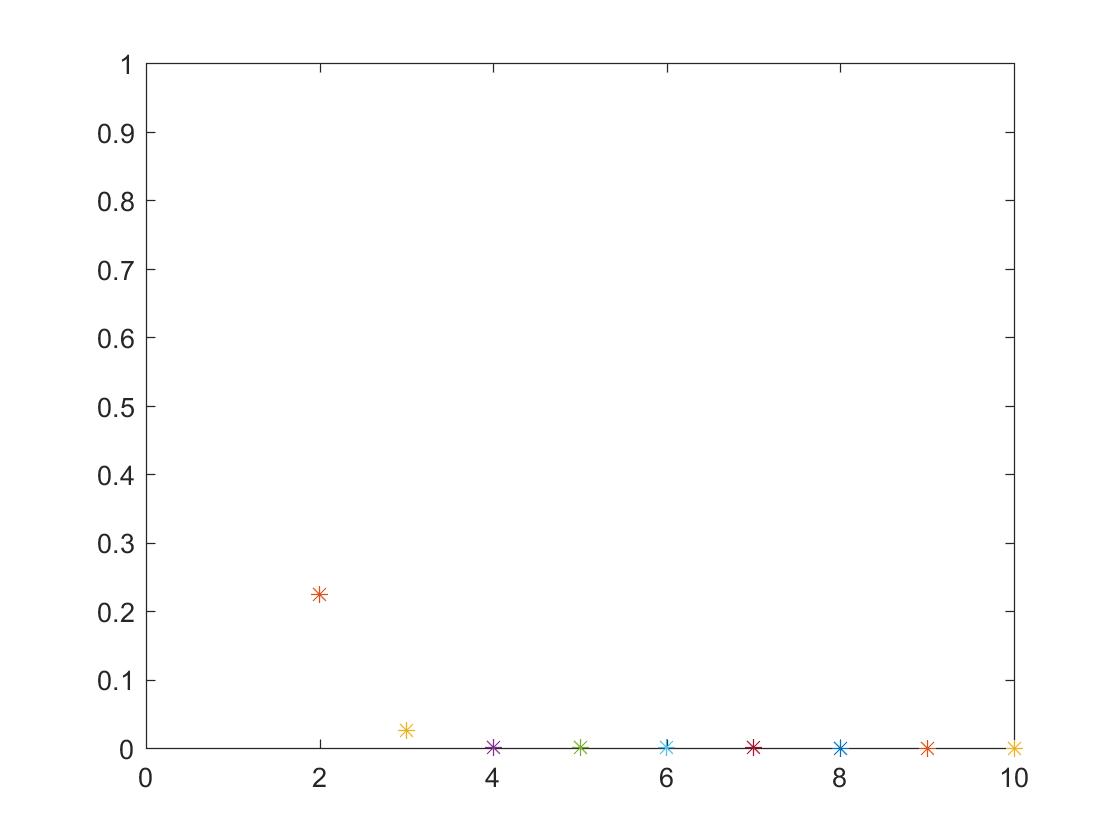}
\caption{Superlinear decrease in error}
\label{fig2}
\end{figure}

Here, we provide some points about the convergence of $\textrm{LTD}$ algorithm:
\begin{enumerate}
\item
Although we do not have convergence proof for $\textrm{LTD}$, our numerical results confirm at leats superlinear convergence. See Figure \ref{fig2}.
\item
As described in Section $2$, in each iteration of $\textrm{LTD}$ algorithm a low-dimension tridiagonal system is needed to improve the signal quality. To determine the most noisy elements, we suggest to select the entries with second difference greater than 70 percent of its maximum value as a rule of thumb. Based on our observtions it is a proper choice in most of the tests.
\item
The values of $kmax$ and $\delta$ depends on data size. In \ref{tab1} proper choices are represented. 
 \end{enumerate}

\begin{table}[h]
\caption{Suggested parametes}\label{tab1}%
\begin{tabular}{@{}ccccc@{}}
\toprule
n & $kmax$ & $\delta$\\
\midrule
100     &   10   &  E-06\\
500     &   10   &  E-05\\
1000   &   100  &  E-04\\
5000   &   100  &  E-04\\
10000 &   200  &  E-03\\
\botrule
\end{tabular}
\end{table}

\section{Numerical Results}\label{sec4}
\begin{figure}[!h]%[!htbp]
\centering
\includegraphics[scale=0.5]{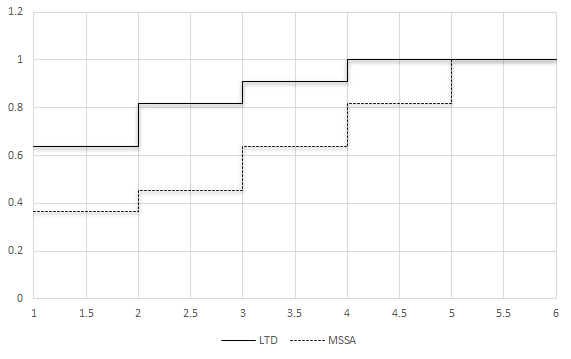}
\caption{Dolan More time profiles for LTD and MSSA algorithms}
\label{fig3}
\end{figure}

In this section, we present the numerical results. We implement LTD and MSSA algorithms in MATLAB 2019b on a computer with a 2.4 GHz core i5 CPU and 8Gb RAM. We then test the codes on real and randomly generated noisy data. To generte random tests, both rand and randn commands are used for the exact data and a normal noise is added. In each experiment, the goal is to capture the added noise as fast as possible. We compare both MSE and time to show the effectiveness of our proposed algorithm in approximating the noise term more precisely and in lower time. In Figure \ref{fig3}, the Dolan More time profiles are shown to verify the fastness of LTD.
Moreover, the average time and MSE are reported for random tests. To provide more accurate results we repeat each experiment 20 times and report the average results in Table \ref{tab2}. As large as the data size is, MSSA tends to outperform our algorithm; however, for data size not greater than 1000 LTD has two desirable features including lower MSE and lower computing time. 

\begin{table}[h]
\caption{Time and MSE comparison for LTD and MSSA \cite{Chen2015}}\label{tab2}%
\begin{tabular}{@{}ccccc@{}}
\toprule
n & LTD time & MSSA time & LTD MSE & MSSA MSE\\
\midrule
100    &  0.2669   &  1.1526  &  0.0813  &  0.1149\\
500    &  0.5703   &  1.2832  &  0.0271  &  0.1198\\
1000  &  1.9634   &  1.4939  &  0.0164  &  0.1231\\
5000  &  4.7623   &  3.6403  &  0.0132  &  0.1238\\
10000 & 21.1465  & 6.1078  &  0.0125  &  0.1246\\
\botrule
\end{tabular}
\end{table}

\section{Concluding Remarks}\label{sec5}
Noise reduction was the target of this paper. The most important contribution was to outline a low computational cost algorithm for detecting small data fluctuations. In our suggested algorithm two phase are introduced: first was to suggest a local tridiagonal model around the most noisy entries to detect the noise and the second was to design a learn/feedback process to decide whether the predicted noise satisfied the necessary quality conditions in next iterations. According to presented numerical results, the presented algorithm was able to detect the small fluctuations with lower men squared error in lower computational time. Working on optimal parallelization techniques can be suggested for future research to denoise large scale data sets.
\bibliography{sn-article}% common bib file
%% if required, the content of .bbl file can be included here once bbl is generated\input{sn-article.bbl}

\end{document}